%% file: main.tex
\documentclass{ecai}
\usepackage{booktabs} 

\newcommand{\includevisio}[2][]{\includegraphics[clip, trim=0.5cm 0.5cm 0.5cm 0.5cm, #1]{#2}} 

\usepackage{array}
\newcommand{\PreserveBackslash}[1]{\let\temp=\\#1\let\\=\temp}
\newcolumntype{C}[1]{>{\PreserveBackslash\centering}p{#1}}
\newcolumntype{R}[1]{>{\PreserveBackslash\raggedleft}p{#1}}
\newcolumntype{L}[1]{>{\PreserveBackslash\raggedright}p{#1}}

\usepackage{todonotes} 
\usepackage{amsmath}
\usepackage{bm}
\usepackage{amsfonts}
\usepackage{amsthm}
\usepackage{multirow}
\usepackage{colortbl}
\usepackage{enumitem}
\usepackage[normalem]{ulem}
\usepackage{times}
\usepackage{graphicx}
\usepackage{latexsym}

\usepackage{xcolor}

\input{dlbook_math}

\begin{document}

\title{Interpretable \& Time-Budget-Constrained Contextualization for Re-Ranking} 

\author{Sebastian Hofst{\"a}tter\institute{TU Wien,
Austria, email: s.hofstaetter@tuwien.ac.at} \and Markus Zlabinger\institute{TU Wien,
Austria, email: markus.zlabinger@tuwien.ac.at} \and Allan Hanbury\institute{TU Wien,
Austria, email: hanbury@ifs.tuwien.ac.at} }

\maketitle
\bibliographystyle{ecai}

\begin{abstract} Search engines operate under a strict time constraint as a fast response is paramount to user satisfaction. Thus, neural re-ranking models have a limited time-budget to re-rank documents. Given the same amount of time, a faster re-ranking model can incorporate more documents than a less efficient one, leading to a higher effectiveness. To utilize this property, we propose TK (Transformer-Kernel): a neural re-ranking model for ad-hoc search using an efficient contextualization mechanism. TK employs a very small number of Transformer layers (up to three) to contextualize query and document word embeddings. To score individual term interactions, we use a document-length enhanced kernel-pooling, which enables users to gain insight into the model. TK offers an optimal ratio between effectiveness and efficiency: under realistic time constraints (max. 200 ms per query) TK achieves the highest effectiveness in comparison to BERT and other re-ranking models. We demonstrate this on three large-scale ranking collections: MSMARCO-Passage, MSMARCO-Document, and TREC CAR. In addition, to gain insight into TK, we perform a clustered query analysis of TK's results, highlighting its strengths and weaknesses on queries with different types of information need and we show how to interpret the cause of ranking differences of two documents by comparing their internal scores.

\end{abstract}
\input{contents/1-introduction.tex}
\input{contents/2-background.tex}
\input{contents/3-model.tex}
\input{contents/4-methodology.tex}

\input{contents/5-res+disc.tex}
\input{contents/6-conclusion.tex}
\vspace{-0.1cm}

\bibliography{my-references}
\end{document}

%% file: dlbook_math.tex
\def\eqref#1{equation~\ref{#1}}

\def\1{\bm{1}}

\DeclareMathAlphabet{\mathsfit}{\encodingdefault}{\sfdefault}{m}{sl}
\SetMathAlphabet{\mathsfit}{bold}{\encodingdefault}{\sfdefault}{bx}{n}



%% file: contents/1-introduction.tex
\vspace{-0.15cm}
\section{Introduction}
\vspace{-0.09cm}

The importance of efficient and fast search engines is well established \cite{kohavi2013online}. Therefore, the time spent on each part of the Information Retrieval (IR) pipeline has to be managed with time-constraints. Naturally, re-ranking models, which improve the effectiveness of initial rankings, need to stay within a certain time-budget to be deployable to a user facing search engine. In recent years neural network based re-ranking models matured and a distinct trade-off emerged between a neural re-ranking model's effectiveness and its efficiency. While IR-specific networks are reasonably fast \cite{Xiong2017, Dai2018, hui2017pacrr}, large Transformer based models \cite{vaswani2017attention}, such as BERT \cite{devlin2018bert}, show substantially better effectiveness at the cost of orders of magnitude longer inference time \cite{Hofstaetter2019_osirrc, macavaney2019, nogueira2019document}. Given the same amount of limited time, a faster re-ranking model can incorporate more documents than a less efficient one, leading to a higher effectiveness.

In this paper, we present TK -- an interpretable neural re-ranking model for ad-hoc retrieval with a focus on a good ratio between efficiency and effectiveness, which is particularly suited for a time-constrained environment. TK is short for Transformer-Kernel -- the two main components of our model (Section \ref{sec:model}). 

TK brings two main contributions in terms of efficiency and explainability. First, we show how a small number of lightweight Transformer layers~\cite{vaswani2017attention} (we evaluate up to three) can effectively contextualize query and document word embeddings. TK's second contribution is a network structure built for explainability. In contrast to BERT-based approaches, we contextualize query and document sequences independent from each other and distill the interactions between terms in a single interaction match matrix, followed by soft-histogram scoring based on kernel-pooling \cite{Xiong2017}. This allows us to explain scoring reasons by probing the model at the point of the information bottleneck to analyze contextualized term representations and interaction patterns.

We conduct experiments on three large retrieval collections: MSMARCO-Passage~\cite{msmarco16}, MSMARCO-Document~\cite{msmarco16}, and TREC CAR 2017~\cite{dietz2017trec}. We evaluate a broad range of traditional and neural ranking models. We introduce time-budget aware evaluation, which varies the re-ranking depth according to the available time and speed of each neural model. Our experiments show that TK is the best model choice for an average re-ranking budget per query under 200 ms for MRR, 500 ms for Recall, and 250 ms for nDCG. At 100 ms per query TK's MRR is 10\% higher, Recall is 40\% higher, and nDCG is 19\% higher than BERT (Section \ref{sec:results}).

To further understand our novel model, we conduct a query-level analysis of TK's effectiveness and explain the cause of ranking differences of two documents. We cluster queries based on their contextualized embeddings and inspect the cluster's median reciprocal rank. This allows us to robustly identify the strengths and weaknesses of our model on different types of the user's information need (Section \ref{sec:query_analysis}). We demonstrate the interpretation capability of the TK model using the scenario in which a user would like to understand, for a given query, why two documents are ranked differently. We visualize word-level similarities (interaction features) and we report intermediate results of important kernels (Section \ref{sec:interpretability}).

We publish the source code of our work at \textit{github.com/sebastian-hofstaetter/transformer-kernel-ranking}. The repository contains all pre-processing and evaluation code, as well as clear and documented neural network implementations using PyTorch~\cite{pytorch2017} and AllenNLP~\cite{Gardner2017AllenNLP}.

In summary, the main contributions of this work are as follows:
\vspace{-0.1cm}
\begin{itemize}
    \item We propose TK: a re-ranking model using contextualized representations for time-constrained applications.
    \begin{itemize}
        \vspace{-0.1cm}
        \item Efficiency: We show that a small number of low-dimensional Transformers contextualize efficiently and effectively.
        \vspace{-0.1cm}
        \item Interpretability: TK's architecture allows to extract and analyze the full information flow at a single point
    \end{itemize}
    \vspace{-0.1cm}
    \item We introduce time-budget \& re-ranking depth aware\\ evaluation of neural IR models.
    \item We conduct a robust query-level analysis and \\demonstrate the interpretability of TK.
\end{itemize}

\newpage

%% file: contents/2-background.tex
\section{Related Work}
\label{sec:background}
\vspace{-0.15cm}
The short history of neural re-ranking models already saw three waves of architectures: representation, interaction, and contextualized interaction models \cite{Guo2016}. The first representation-focused neural IR models unsuccessfully tried to match single vector representations per query and document \cite{mitra2018introduction}. Then, interaction-focused models moved to a more fine-grained modelling of query-document interactions based on a match-matrix. Now, contextualization in various forms offers the most effective approaches. 

The core of interaction approaches are term by term similarities. A key success factor are fine-tuned word representations, covering most of the indexed vocabulary \cite{Hofstaetter2019_sigir}. Various approaches exist to reduce the match-matrix of term similarities to the matching score: using stacked Convolutional Neural Networks (CNN) \cite{Pang2016, mitra2019updated}, parallel single-layered CNNs for n-gram interaction modelling \cite{hui2017pacrr}, recurrent neural networks \cite{Fan2018}, and position independent counting methods. Guo et al. \cite{Guo2016} showed the promise of counting interactions with the histogram-based DRMM model. However, it suffered from the non-differentability of a hard histogram method and the resulting lack of fine-tuned word representations. Xiong et al.~\cite{Xiong2017} improve on the idea and propose the kernel-pooling technique as part of the KNRM model. Conceptually, it approximates a histogram with a set of Gaussian kernel functions for different similarity ranges instead of a hard binning. The kernel-pooling offers a solid foundation for analysis and interpretability \cite{Pyreddy2018}, whereas pattern-based methods are harder to interpret post-hoc \cite{Zeon2019}.

Contextualization allows neural IR models to vary the importance of otherwise identical term matches. The neural CO-PACRR model \cite{hui2018co} provides a lightweight contextualization. It averages word vectors with a sliding window and appends their similarities to the non-contextualized similarities of the PACRR \cite{hui2017pacrr} model. The CONV-KNRM model~\cite{Dai2018} extends KNRM by adding a CNN layer on top of the word embeddings, enabling word-level n-gram representation learning -- a local contextualization, fixed by the n-gram size hyperparameter. 

Vaswani et al. \cite{vaswani2017attention} proposed the Transformer architecture in the context of language translation. Their encoder-decoder is built of Transformer layers, each containing multi-head self attention. These Transformer layers are the building blocks of versatile multi-task architectures, such as BERT  \cite{devlin2018bert} and XLNet \cite{yang2019xlnet}. These models rely on a computationally intensive pre-training. Publicly available pre-trained models can then be fine-tuned for various tasks, including pairwise sequence classification. Nogueira et al. \cite{nogueira2019passage,nogueira2019document} first showed the applicability of BERT for re-ranking and the resulting substantial effectiveness gains. MacAvaney et al. \cite{macavaney2019} show that it is beneficial to combine BERT's classification label with the output of interaction-based neural IR models. Both note that using BERT comes at a substantial performance cost -- BERT taking two orders of magnitude longer than a simple word embedding.

In traditional learning-to-rank the trade-off between effectiveness and efficiency has been thoroughly studied  \cite{wang2010learning,wang2010ranking,xu2012greedy,capannini2016quality}. This includes applying a temporal constraint on the number of features that are selected for a re-ranking model \cite{wang2010ranking}, incorporating an efficiency metric in the training of linear rankers \cite{wang2010learning}, and comparing the effectiveness and efficiency of various learning-to-rank algorithms \cite{capannini2016quality}. In web search the speed of a response is crucial as determined by Kohavi et al. \cite{kohavi2013online} in a large scale experiment, however, in some expert tasks, users are willing to wait longer for better results, so that the best model choice becomes task dependent \cite{teevan2013slow}.

Recently, the issue of efficiency gained traction in the neural IR community. Hofst\"atter et al. \cite{Hofstaetter2019_osirrc} establish efficiency baselines for common neural IR models (including BERT) and propose to incorporate speed metrics in replicability campaigns and public leaderboards. One way to make neural IR models faster at query time is to offload computation to the indexing phase, either by assuming query term independence \cite{mitra2019incorporating} or by approximating interaction similarities \cite{ji2019efficient}. When a large number of pre-trained Transformer layers is involved, inference can be sped up by removing later layers and scoring the intermediate results instead \cite{macavaney2019} or by pruning unnecessary attention-heads \cite{voita2019analyzing}. In this work we speed up Transformer contextualization by using very small and few Transformer blocks.

%% file: contents/3-model.tex
\section{TK: Transformer-Kernel Model}
\label{sec:model}

\begin{figure*}
    \centering
    \includevisio[width=1\textwidth]{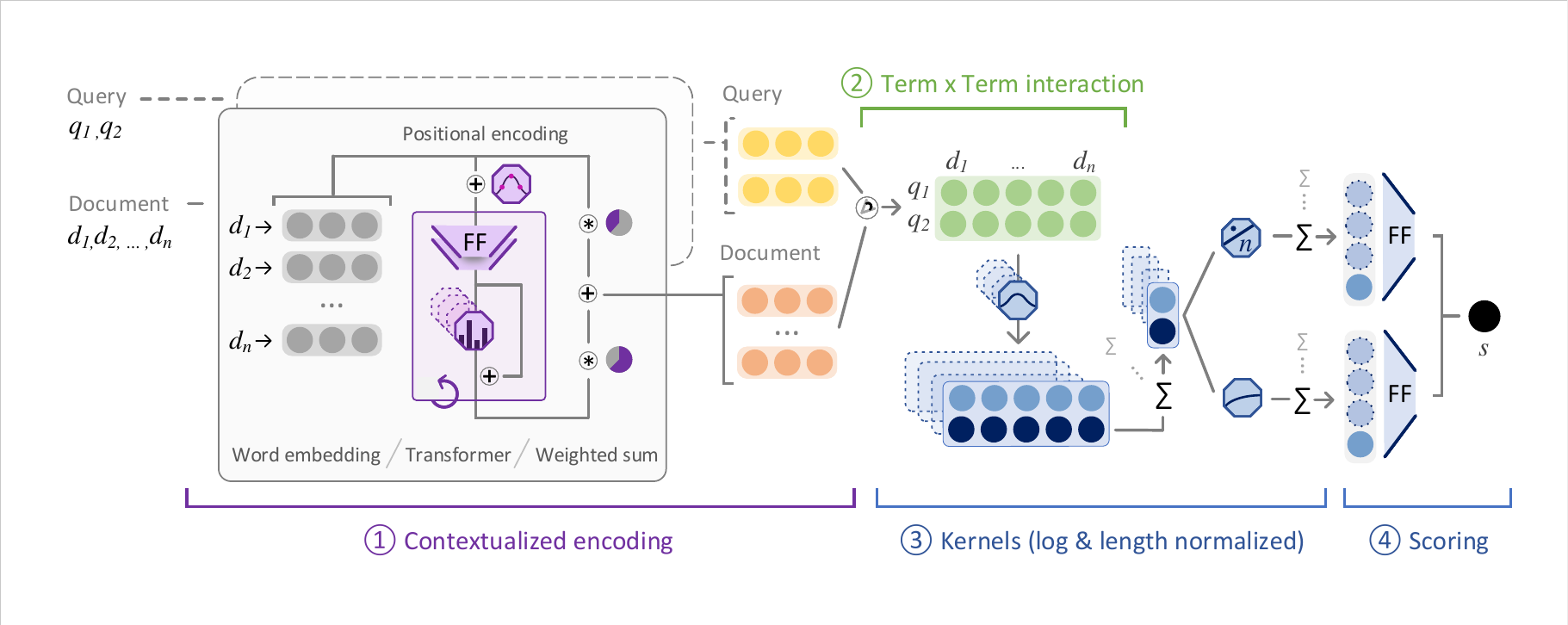}
    \vspace{-0.8cm}
    \caption{The TK model architecture: \raisebox{.5pt}{\textcircled{\raisebox{-.85pt} {1}}}
 We contextualize query and document sequences individually. \raisebox{.5pt}{\textcircled{\raisebox{-.85pt} {2}}} The interaction match-matrix is created with pairwise cosine similarities. \raisebox{.5pt}{\textcircled{\raisebox{-.85pt} {3}}} Each kernel creates a new feature matrix. Then, the document dimension is summed and we normalize each query-term feature by logarithm and document length. \raisebox{.5pt}{\textcircled{\raisebox{-.85pt} {4}}} We combine log- and length-normalized scores with a single feed-forward (FF) layer to form the final result score.}
    \label{fig:tk_model_architecture}
\end{figure*}

In this section, we present TK, our Transformer-Kernel neural re-ranking model. In the following, we describe how we learn contextualized term representations (Section \ref{sec:tk_context}) and how we transparently score their interactions (Section \ref{sec:tk_scoring}). Figure \ref{fig:tk_model_architecture} gives an overview of the TK architecture.

\subsection{Contextualized Term Representation}
\label{sec:tk_context}

TK uses a hybrid contextualization approach. The base representations are single-vector-per-word embeddings \cite{pennington2014glove}. We chose a simple word embedding structure over more complex methods -- such as FastText \cite{bojanowski2017enriching} or ELMo \cite{Peters2018} -- as it offers many benefits in practice. Word embeddings are easy to pre-train on domain specific data \cite{Hofstaetter2019_ecir}. They require only one id per term, making the index consume less disk space, once prepared for re-ranking. Most importantly, at query time, their selection is a fast memory lookup.

In the contextualization phase of the TK model, we process query ${q}_{1:m}$ and document sequences ${d}_{1:n}$ separately, however the learned parameters are shared. The input consists of two sequences of query and document ids. We employ the lookup based word embedding to select non-contextualized representations for each term. The hybrid-contextualized representation $\hat{t}_i$ of a term with word embedding $t_i$ over its whole input sequence $t_{1:n}$ is defined as:
\begin{equation}
\hat{t}_i = t_i * \alpha + \operatorname{context}(t_{1:n})_i * (1 - \alpha)
\end{equation}

We regulate the influence of the contextualization by the end-to-end learned $\alpha$ parameter. This allows the model to decide the intensity of the contextualization. We calculate the $\operatorname{context}(t_{1:n})$ with a set of Transformer layers \cite{vaswani2017attention}. First, the input sequence is fused with a positional encoding to form $p_{1:n}$, followed by a set of $l$ Transformer layers:  
\begin{equation}
\operatorname{Transformer_l}(p_{1:n}) = \operatorname{MultiHead}(\operatorname{FF}(p_{1:n})) + \operatorname{FF}(p_{1:n})
\end{equation}

Here, $\operatorname{FF}$ is a two-layer fully connected feed-forward layer including a non-linear activation function. The $MultiHead$ module projects the input sequence (via $W_{i}^{*}$) to query, key, and value inputs of the scaled dot-product attention for each attention head. Then the results of the attention heads are concatenated and projected to the output (via $W^{O}$):  
\begin{equation}
\begin{aligned} 
\operatorname{MultiHead}(p_{1:n}) &=\operatorname{Concat}(head_{1}, ..., head_{h}) W^{O} \\
\text{where } head_i &= \operatorname{softmax}\left(\frac{(p_{1:n} W_{i}^{Q}) (p_{1:n} W_{i}^{K})^{T}}{\sqrt{d_{k}}}\right) (p_{1:n} W_{i}^{V})
\end{aligned}
\end{equation}

We select Transformers for contextualization, because their positional encoding and sequence wide self-attention allows for local and global contextualization at the same time. This makes TK more powerful than previous local-only contextualization methods used in CONV-KNRM \cite{Dai2018} and CO-PACRR \cite{hui2018co}.

\subsection{Interaction Scoring}
\label{sec:tk_scoring}

After the contextualization, we match the query sequence $\hat{q}_{1:m}$ and document sequence $\hat{d}_{1:n}$ together in a single match-matrix $M~\in~\mathbb{R}^{q_{\text{len}} \times  d_{\text{len}}}$ with pairwise cosine similarity as interaction extractor:
\begin{equation}
M_{i,j} = \cos(\hat{q_i},\hat{d_j})
\end{equation}

Then, we transform each entry in $M$ with a set of RBF-kernels \cite{Xiong2017}. Each kernel focuses on a specific similarity range with center $\mu_{k}$. The size of all ranges is set by $\sigma$. In contrast to Xiong et al. \cite{Xiong2017} we do not employ an exact match kernel -- as contextualized representations do not produce exact matches. Each kernel results in a matrix $K \in \mathbb{R}^{q_{\text{len}} \times  d_{\text{len}}}$: 
\begin{equation}
K^{k}_{i,j} = \exp \left(-\frac{\left(M_{i j}-\mu_{k}\right)^{2}}{2 \sigma^{2}}\right)
\end{equation}

Now, we process each kernel matrix in parallel, and we begin by summing the document dimension $j$ for each query term and kernel:
\begin{equation}
K^{k}_{i} = \sum_{j} K^{k}_{i,j}
\end{equation}

At this point -- as shown in Figure \ref{fig:tk_model_architecture} -- the model flow splits into two paths: log normalization and length normalization. The log normalization applies a logarithm with base $b$ to each query term before summing them up:
\begin{equation}
s^{k}_{\text{log}} = \sum_{i} \log_b\left( K^{k}_{i} \right)
\end{equation}

To incorporate the notion of different document lengths into the model we enhance the pooling process with document length normalization. We dampen the magnitude of each query term signal by the document length:
\begin{equation}
s^{k}_{\text{len}} = \sum_{i} \frac{K^{k}_{i}}{d_{\text{len}}}
\end{equation}

Now, the set of kernel scores (one value per kernel) is weighted and summed up with a simple linear layer ($W_{log},W_{len})$ to produce a scalar, for both the log-normalized and length normalized kernels: 
\begin{align}
s_{\text{log}} = s^{k}_{\text{log}} W_{log} && s_{\text{len}} = s^{k}_{\text{len}} W_{len}
\end{align}

Finally, we compute the final score of the query-document pair as a weighted sum of the log-normalized and the length-normalized scores:
\begin{equation}
s = s_{\text{log}} * \beta + s_{\text{len}} * \gamma 
\end{equation}

We employ kernel-pooling, because it makes inspecting temporary scoring results more feasible compared to pattern based scoring methods (for example PACRR \cite{hui2017pacrr}). Each kernel is applied to the full document and the row-wise and the column-wise summing of the match-matrix allow to inspect individual matches independent from each other.
\subsection{Difference to Related Work}

The main differences of TK in comparison to BERT \cite{nogueira2019passage} are:
\begin{itemize}
    \item TK's contextualization uses fewer and lower dimensional Transformer layers with less attention heads. This makes the query-time inference of TK with 2 layers 40 times faster than BERT-Base with 12 layers.
    \item TK contextualizes query and document sequences independently; each contextualized term is represented by a single vector (available for analysis). BERT operates on a concatenated sequence of the query and the document, entangling the representations in each layer.
    \item The network structure of TK makes it possible to analyze the model for interpretability and further studies. TK has an \textit{information bottleneck} built in, through which all term information is distilled: the query and document term interactions happen in a single match matrix, containing exactly one cosine similarity value for each term pair. BERT on the other hand has a continuous stream of interactions in each layer and each attention head, making a focused analysis unfeasible.
\end{itemize}

\noindent The differences of TK to previous kernel-pooling methods are:

\begin{itemize}
    \item KNRM \cite{Xiong2017} uses only word embeddings, therefore a match does not have context or positional information.
    \item CONV-KNRM \cite{Dai2018} uses a local-contextualization with limited positional information in the form of n-gram learning with CNNs. It cross-matches all n-grams in $n^2$ match matrices, reducing the analyzability.
\end{itemize}

%% file: contents/4-methodology.tex
\section{Experiment Setup}
\label{sec:experiment}
\vspace{-0.1cm}

For the first stage indexing and retrieval we use the Anserini toolkit \cite{Yang2017} to compute baselines as well as the initial ranking lists, which we use to generate training and evaluation inputs for the neural models. For our neural re-ranking training and inference we use PyTorch~\cite{pytorch2017} and AllenNLP~\cite{Gardner2017AllenNLP}. For BERT support we use the pytorch-transformer library\footnote{https://github.com/huggingface/pytorch-transformers}. 
We train all neural models with a pairwise hinge loss. We conduct our experiments and report timings with an NVIDIA GTX 1080 TI (11GB memory) GPU. 

\subsection{Baselines}
\vspace{-0.1cm}
We use tuned BM25, Language Modelling with Dirichlet smoothing (LM), and RM3 \cite{Abdul-Jaleel2004rm3} as traditional retrieval method baselines.  In the following we give an overview over our neural baselines: 

\textbf{\emph{MatchPyramid}}~\cite{Pang2016} applies several stacked CNN layers with max-pooling on top of a term-by-term interaction matrix. The pooling sizes become smaller with each layer -- like a pyramid. 

\textbf{\emph{DUET}}~\cite{mitra2019updated} is a hybrid model which applies CNNs to local term-by-term interactions and it learns a single representation for query and document and then measures the similarity between the two vectors. The two paths are combined an jointly scored. 

\textbf{\emph{PACRR}}~\cite{hui2017pacrr} applies different sized CNNs on the match matrix followed by a max pooling. In contrast to MatchPyramid, the single CNN layer focuses on different n-gram sizes. 

\textbf{\emph{CO-PACRR}}~\cite{hui2018co} extends the PACRR model with additional contextualized similarities (via fixed window neighborhood mean vectors) and improves the robustness of PACRR's pooling strategy with randomization during training. 

\textbf{\emph{KNRM}}~\cite{Xiong2017} uses a soft-histogram (differentiable Gaussian kernel functions) on top of the interaction matrix of query and document embeddings -- summing the interactions by their similarity. 

\textbf{\emph{CONV-KNRM}}~\cite{Dai2018} applies a CNN over the query and document word embeddings, resulting in word-level n-gram representations. CONV-KNRM cross-matches $n$-grams and subsequently scores the interactions with $n$ KNRM instances. 

\textbf{\emph{BERT$_{\textbf{[CLS]}}$}}~\cite{devlin2018bert} is a multi-task Transformer based NLP model. Pre-trained instances are commonly available in large and small sizes -- we experiment with both. We follow Nogueira et al. \cite{nogueira2019passage} and first concatenate the query and document sequences. To score the pair, we use the representation of BERT's \textit{[CLS]} token and a linear layer.

\vspace{-0.1cm}
\subsection{Datasets \& Resources}
\vspace{-0.1cm}

We train and evaluate our models on three web-focused test collections. The size statistics are shown in Table \ref{tab:collection_stats} and in the following we describe the collections in more detail:

\textbf{\emph{MSMARCO}}~\cite{msmarco16} collections are based on real Bing queries and results. We use both the \textit{Passage} and the \textit{Document} version with different sets of queries. Originally purposed for the question answering task, the annotation data is now used to provide ranking labels for retrieval results\footnote{https://github.com/microsoft/MSMARCO-Passage-Ranking}. If a passage contains the answer to a query (judged by a human annotator) it is deemed relevant in the retrieval task as well as the document containing it.

\textbf{\emph{TREC CAR}}~\cite{dietz2017trec} is created as part of the TREC Complex Answer Retrieval (CAR) task in 2017. It is based on Wikipedia sections: the heading is used as the query and the section body is the deemed relevant paragraph in the automatic annotations.

In addition to the collections, we use pre-trained GloVe~\cite{pennington2014glove} word embeddings with 300 dimensions\footnote{42B CommonCrawl lower-cased: \textit{https://nlp.stanford.edu/projects/glove/}} for all non-BERT neural models and pre-trained weights for BERT from pytorch-transformer.

\begin{table}
    \centering
    \vspace{-0.3cm}
    \caption{Collection statistics}
    \label{tab:collection_stats}
    \begin{tabular}{lrr!{\color{lightgray}\vrule}rr}
       \toprule
       \multirow{2}{*}{\textbf{Collection}} & 
       \multirow{2}{*}{\textbf{\# Docs.}} &
       \multirow{2}{*}{\textbf{\# Terms}} &
    \multicolumn{2}{c}{\textbf{\# Queries}} \\
       & && \textbf{Val.} & \textbf{Test} \\ \midrule
       \textbf{MS-Passage}  & 8,841,823   & 1,834,055 &  6,980 & 48,598  \\
       \textbf{MS-Document} & 3,213,835   & 44,991,958 & 5,000 & 5,193  \\
       \textbf{TREC CAR}    & 29,794,697  & 6,682,592  & 5,000 & 2,254  \\
        \bottomrule
    \end{tabular}
\end{table}

\input{contents/res-table-main.tex}
\vspace{-0.1cm}
\subsection{Parameter Settings}
\vspace{-0.1cm}

We cap the query length at $30$ tokens and the document length at $200$ tokens. For MSMARCO-Passage and TREC CAR this only removes a modest amount of outliers, however, for the MSMARCO-Document collection a majority of documents is longer than 200 tokens. Increasing the cap to fully include most documents would render all evaluated neural IR models less effective or unfeasible for efficiency reasons. Addressing this issue is out of scope of this work, although we plan to address it in future work.
We use the Adam \cite{kingma2014adam} optimizer with a learning rate of $10^{-4}$ for word embeddings and contextualization layers, $10^{-3}$ for all other non-BERT network layers, and $10^{-6}$ for BERT fine-tuning. We employ early stopping, based on the best MRR@10 value of the validation set. We use a training batch size of 64. For evaluation we use a batch size of 256 for all non-BERT models and a batch size of 4 for BERT. We use a vocabulary of all terms with a minimum collection occurrence of 5 for MSMARCO-Passage and TREC CAR; and a collection minimum of 10 for MSMARCO-Document as it contains more unique terms.

Regarding model-specific parameters, for the Transformer layers in TK we evaluate 1, 2, and 3 layers, each with 16 attention heads with size 32 and a feed-forward dimension of 100. For log-normalization in TK we use a base of 2. For kernel-pooling (in TK, KNRM, CONV-KNRM) we set the number of kernels to $11$ with the mean values of the Gaussian kernels varying from $-1$ to $+1$, and standard deviation of $0.1$ for all kernels (KNRM \& CONV-KNRM use $0.0001$ for exact matching on the first kernel). CONV-KNRM's n-gram size is set to 3 and the CNN features are set to $128$. In the MatchPyramid model, we set the number of CNN layers to $5$, each with kernel size $3\times3$ and 16 convolution channels. We use DUET without dropout and a document pooling width of 100. For PACRR and CO-PACRR we use a maximum n-gram size of 3, 32 CNN features, and a k-max pooling of 5. For the traditional retrieval models we use the tuned parameters from the Anserini documentation.

%% file: contents/res-table-main.tex
\begin{table*}[t!]
    \centering
    \caption{Unconstrained effectiveness results on the test sets. Each measure is using a cutoff of 10. \textit{Depth} refers to the re-ranking depth (which is tuned on MRR@10 on the validation set and the number shown per model is applied on the test set)}
    \label{tab:all_results}
    \setlength\tabcolsep{5pt}
    \small
    \begin{tabular}{l!{\color{lightgray}\vrule}rrrr!{\color{lightgray}\vrule}rrrr!{\color{lightgray}\vrule}rrrr!{\color{lightgray}\vrule}r}
       \toprule
       \multirow{2}{*}{\textbf{Model}}&
       \multicolumn{4}{c!{\color{lightgray}\vrule}}{\textbf{MSMARCO-Passage}}&
       \multicolumn{4}{c!{\color{lightgray}\vrule}}{\textbf{MSMARCO-Document}}&
       \multicolumn{4}{c!{\color{lightgray}\vrule}}{\textbf{TREC CAR}} & \textbf{Average}\\
       & MRR & Recall & nDCG & Depth & MRR & Recall & nDCG & Depth & MRR & Recall & nDCG & Depth & \textbf{Docs./ms}\\
        \midrule
        \textit{\textbf{BM25}}    & \textbf{0.194} & \textbf{0.402} &\textbf{ 0.241 }& - & \textbf{0.252} & \textbf{0.500 }& \textbf{0.311} & - & \textbf{0.221} & \textbf{0.259} & \textbf{0.190} & - & - \\
        \textit{\textbf{LM}}      & 0.171 & 0.358 & 0.213 & - & 0.202 & 0.423 & 0.254 & - & 0.190 & 0.222 & 0.166 & - & - \\
        \textit{\textbf{RM3}}     & 0.169 & 0.388 & 0.219 & - & 0.156 & 0.367 & 0.206 & - & 0.220 & 0.253 & 0.189 & - & - \\
        \midrule
        
        \textit{\textbf{MatchPyramid}}     & 0.249 & 0.476 & 0.301 & 71  & 0.286 & 0.531 & 0.344 & 15 & 0.238 & 0.279 & 0.205 & 40 & 27 \\
        \textit{\textbf{DUET}}             & 0.248 & 0.468 & 0.299 & 42  & 0.266 & 0.520 & 0.327 & 15 & 0.233 & 0.272 & 0.199 & 39 & 14 \\
        \textit{\textbf{PACRR}}            & 0.259 & 0.493 & 0.313 & 619 & 0.283 & 0.536 & 0.344 & 15 & 0.210 & 0.257 & 0.181 & 24 & 22 \\
        \textit{\textbf{CO-PACRR}}         & 0.273 & 0.514 & 0.328 & 987 & 0.284 & 0.543 & 0.345 & 19 & 0.224 & 0.267 & 0.193 & 23 & 14 \\

        \textit{\textbf{KNRM}}             & 0.235 & 0.465 & 0.288 & 127 & 0.261 & 0.519 & 0.323 & 14 & 0.191 & 0.213 & 0.163 & 6 & \textbf{49} \\
        \textit{\textbf{CONV-KNRM}}        & 0.277 & 0.519 & 0.332 & 967 & 0.283 & 0.542 & 0.345 & 19 & 0.223 & 0.275 & 0.194 & 30 & 10 \\

        \textit{\textbf{BERT-Base}}        & \textbf{0.376} & \textbf{0.639} & \textbf{0.437} & 997 & \textbf{0.352} & 0.623 & \textbf{0.417} & 58 & 0.388 & 0.426 & 0.333 & 650 & 0.1 \\
        \textit{\textbf{BERT-Large}}       & 0.366 & 0.627 & 0.426 & 997 & 0.350 & \textbf{0.630} & \textbf{0.417} & 93 & \textbf{0.444} & \textbf{0.475} & \textbf{0.385} & 650 & 0.03 \\
        \midrule
        \textit{\textbf{TK -- 1 Layer}}    & 0.303 & 0.560 & 0.361 & 997 & 0.305 & 0.572 & 0.369 & 29 & 0.285 & 0.312 & 0.241 & 63 & \textbf{6} \\
        \textit{\textbf{TK -- 2 Layer}}    & 0.311 & 0.564 & 0.369 & 997 & 0.312 & 0.577 & 0.375 & 29 & 0.305 & \textbf{0.329} & 0.258 & 86 & 4 \\
        \textit{\textbf{TK -- 3 Layer}}    & \textbf{0.314} & \textbf{0.570} & \textbf{0.373} & 997 & \textbf{0.316} & \textbf{0.586} & \textbf{0.380} & 31 & \textbf{0.307} & 0.328 & \textbf{0.259} & 72 & 2 \\
        \bottomrule
    \end{tabular}
    \vspace{-0.1cm}
\end{table*}

%% file: contents/5-res+disc.tex
\begin{figure*}[t]
    \centering
    \begin{minipage}{.30\textwidth}
    \includegraphics[clip,trim={0.5cm 0.5cm 0.5cm 0.5cm} ,width=\textwidth]{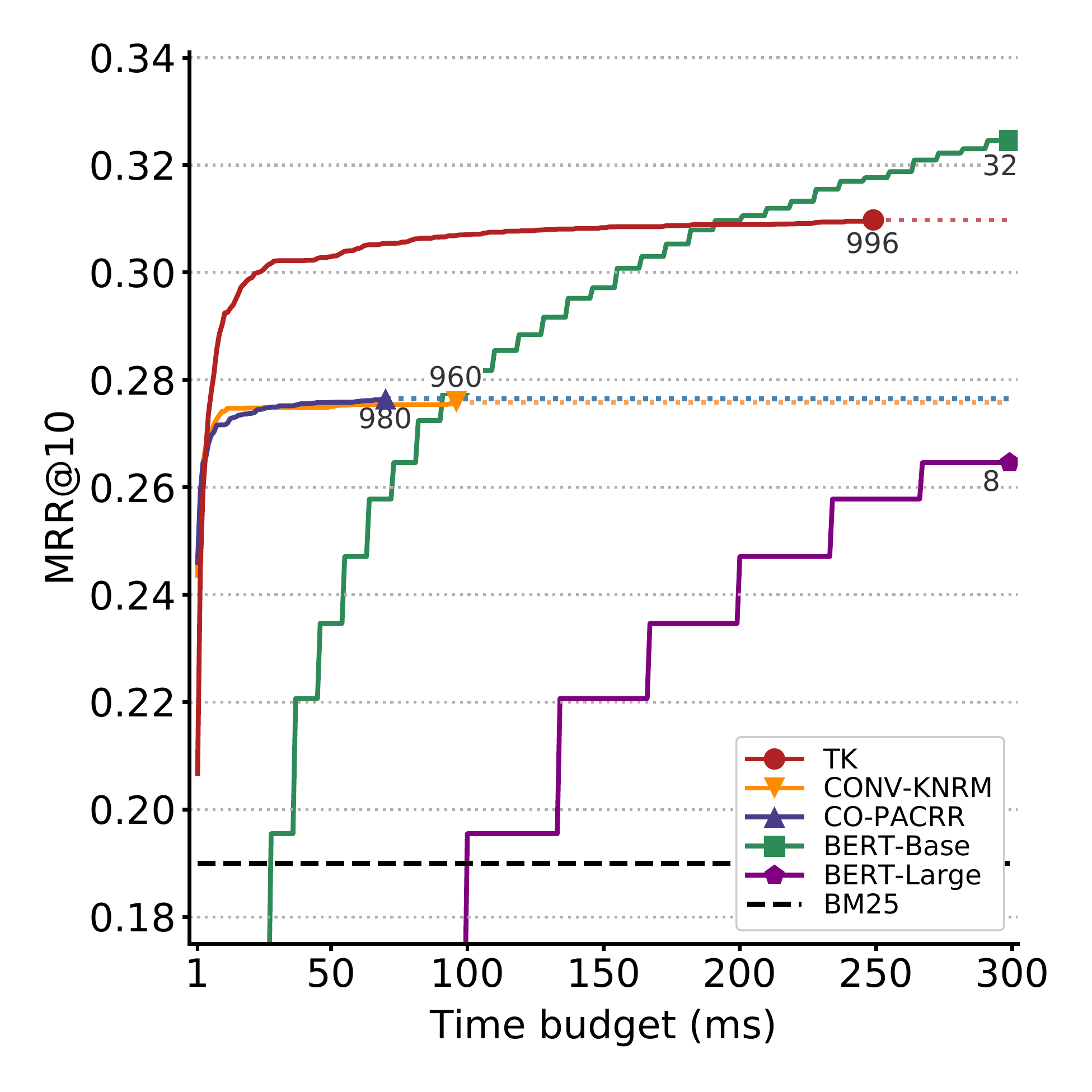}
    \end{minipage}%
    \hfill
    \begin{minipage}{.30\textwidth}
    \includegraphics[clip,trim={0.5cm 0.5cm 0.5cm 0.5cm} ,width=\textwidth]{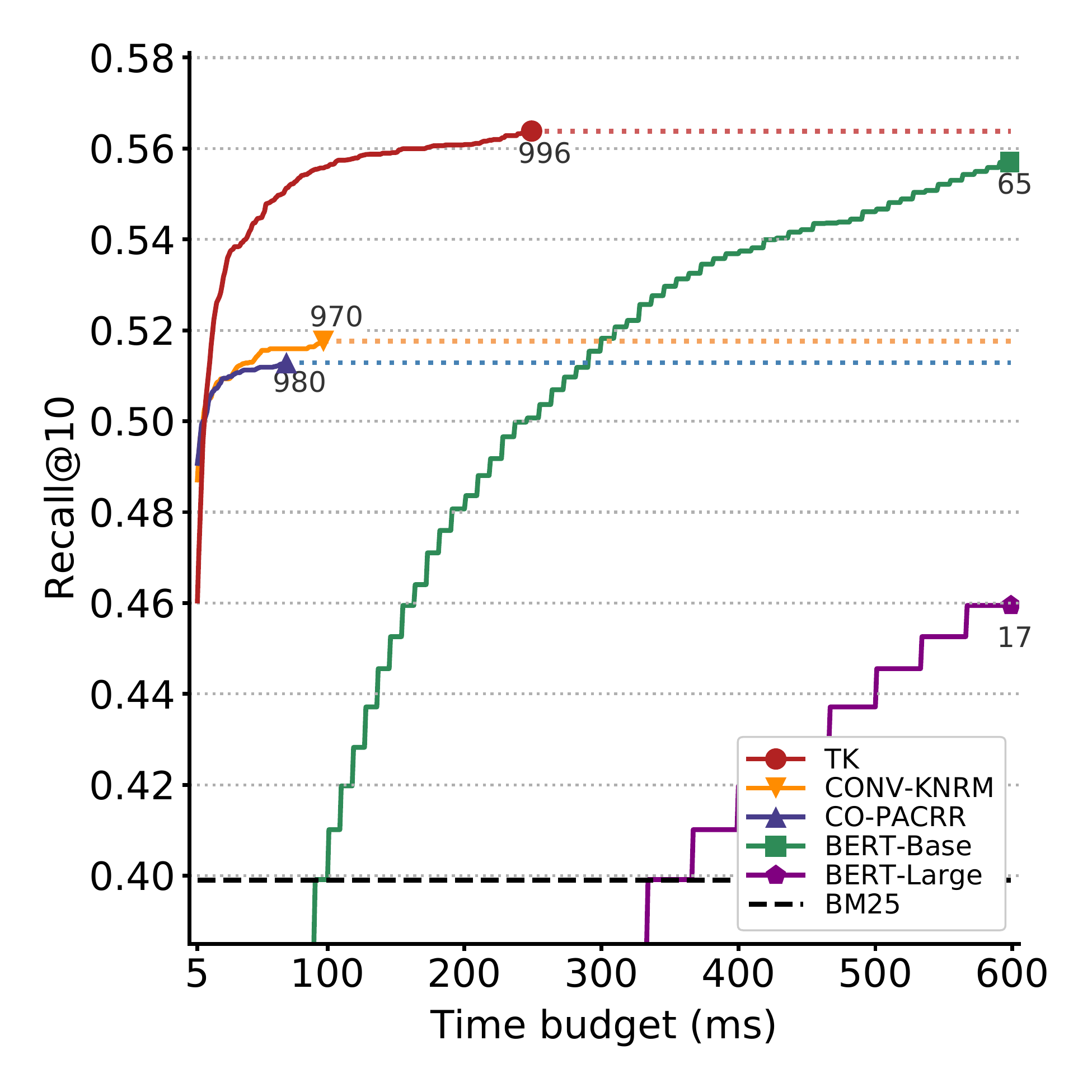}
    \end{minipage}%
    \hfill
    \begin{minipage}{.30\textwidth}
    \includegraphics[clip,trim={0.5cm 0.5cm 0.5cm 0.5cm} ,width=\textwidth]{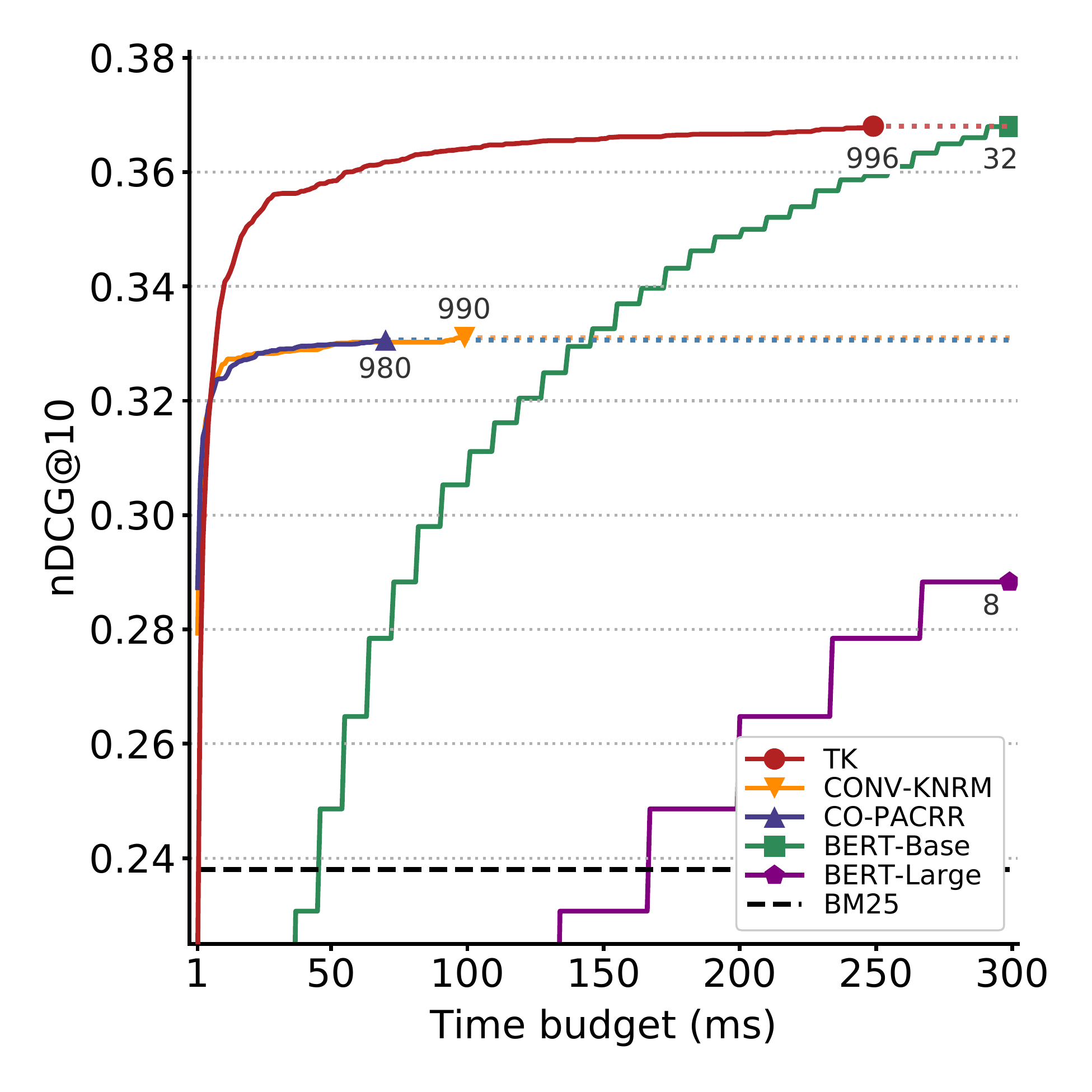}
    \end{minipage}
    \begin{minipage}{\textwidth}{\begin{center} \textbf{(a) MSMARCO-Passage} \end{center}}\end{minipage}
    \begin{minipage}{.30\textwidth}
    \includegraphics[clip,trim={0.5cm 0.5cm 0.5cm 0.5cm} ,width=\textwidth]{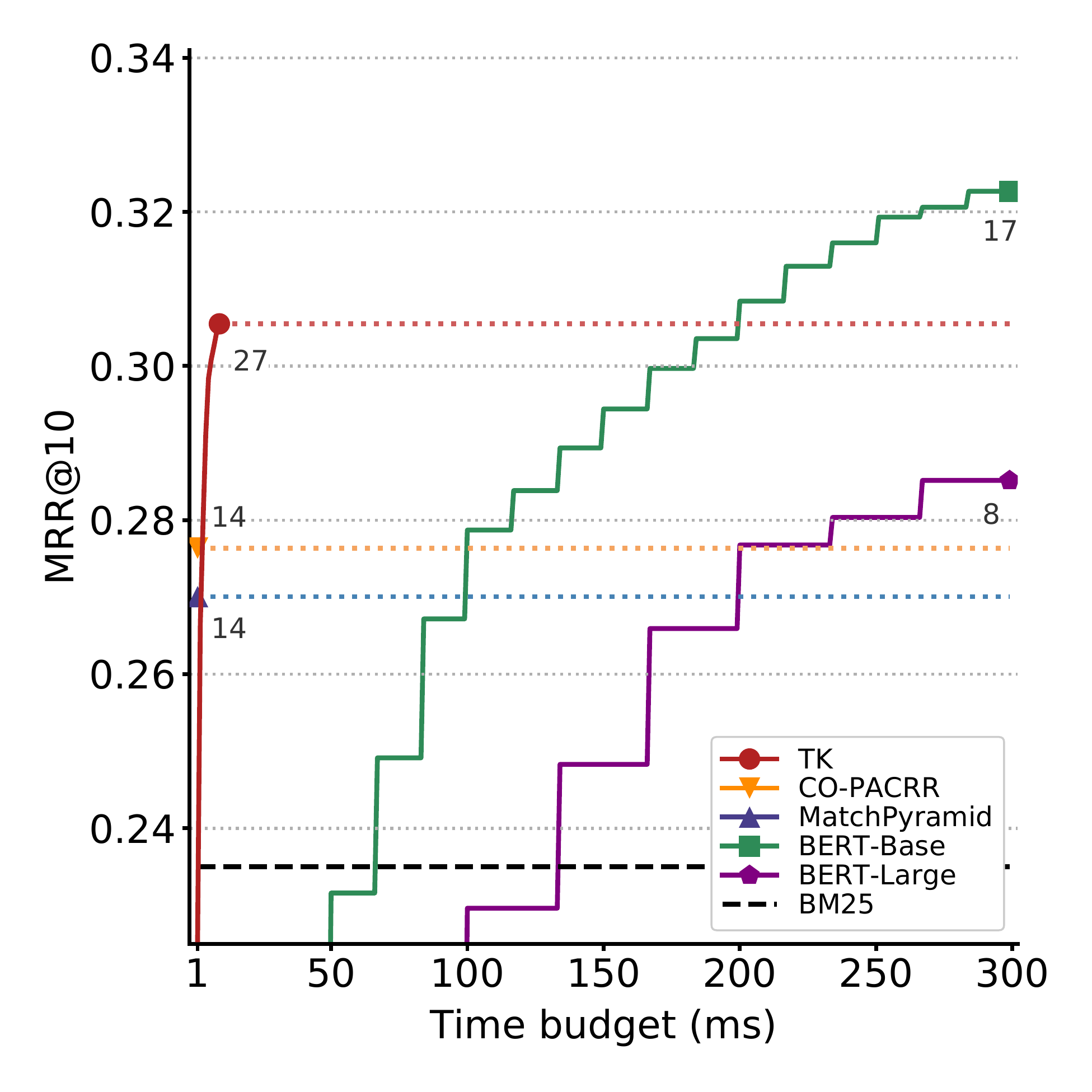}
    \end{minipage}%
    \hfill
    \begin{minipage}{.30\textwidth}
    \includegraphics[clip,trim={0.5cm 0.5cm 0.5cm 0.5cm} ,width=\textwidth]{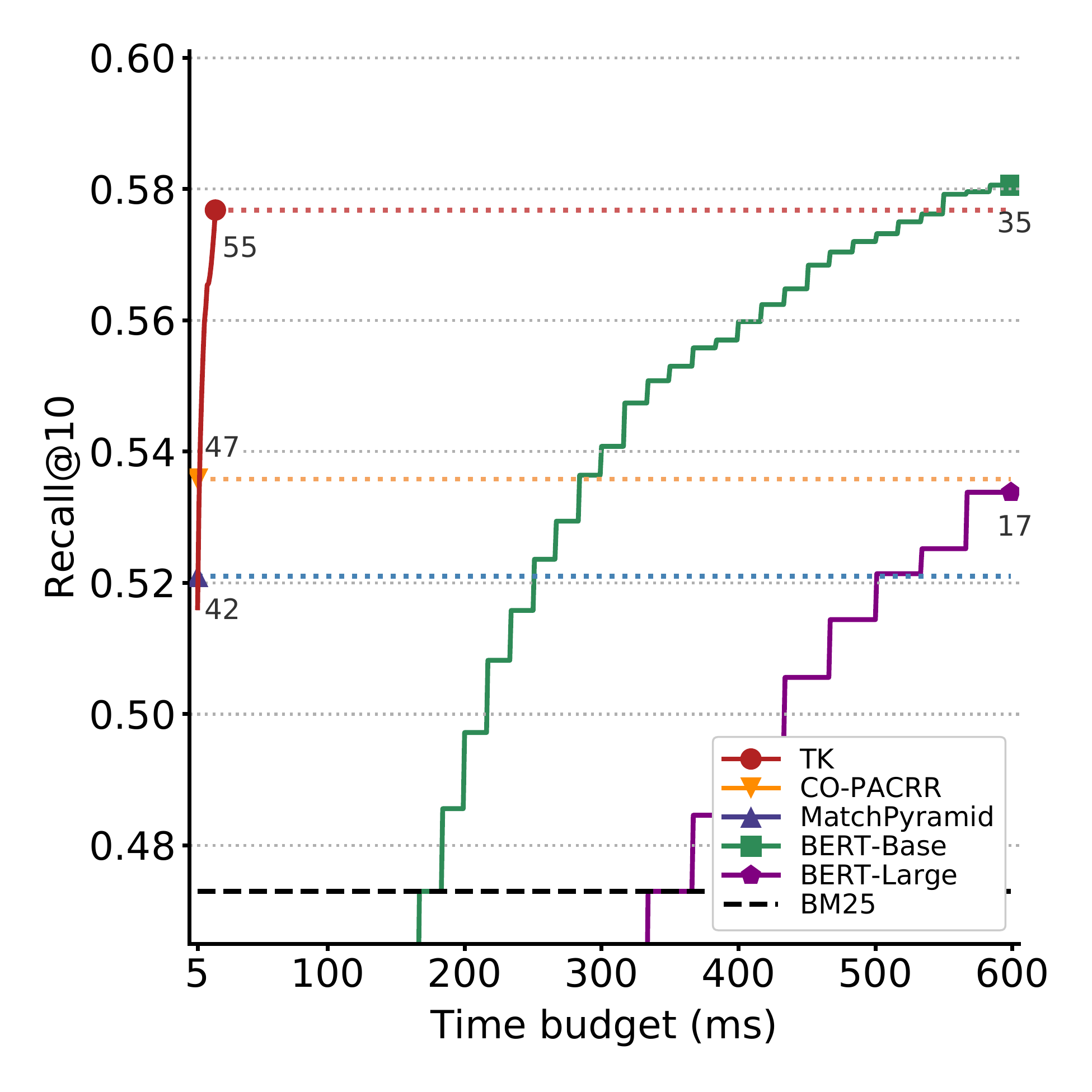}
    \end{minipage}%
    \hfill
    \begin{minipage}{.30\textwidth}
    \includegraphics[clip,trim={0.5cm 0.5cm 0.5cm 0.5cm} ,width=\textwidth]{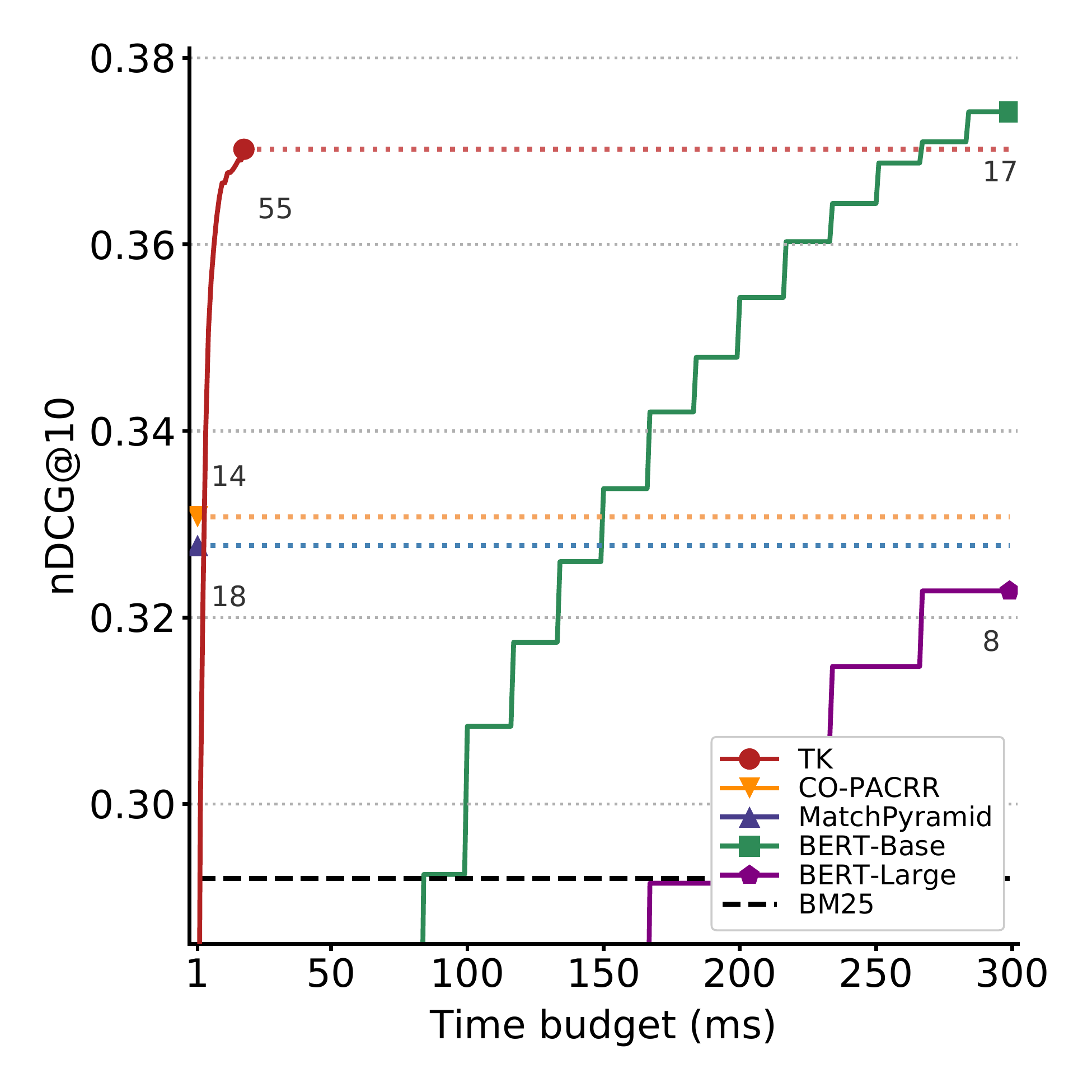}
    \end{minipage}
    
    \begin{minipage}{\textwidth}{ \begin{center}\textbf{(b) MSMARCO-Document}\end{center} }\end{minipage}
    \begin{minipage}{.30\textwidth}
    \includegraphics[clip,trim={0.5cm 0.5cm 0.5cm 0.5cm} ,width=\textwidth]{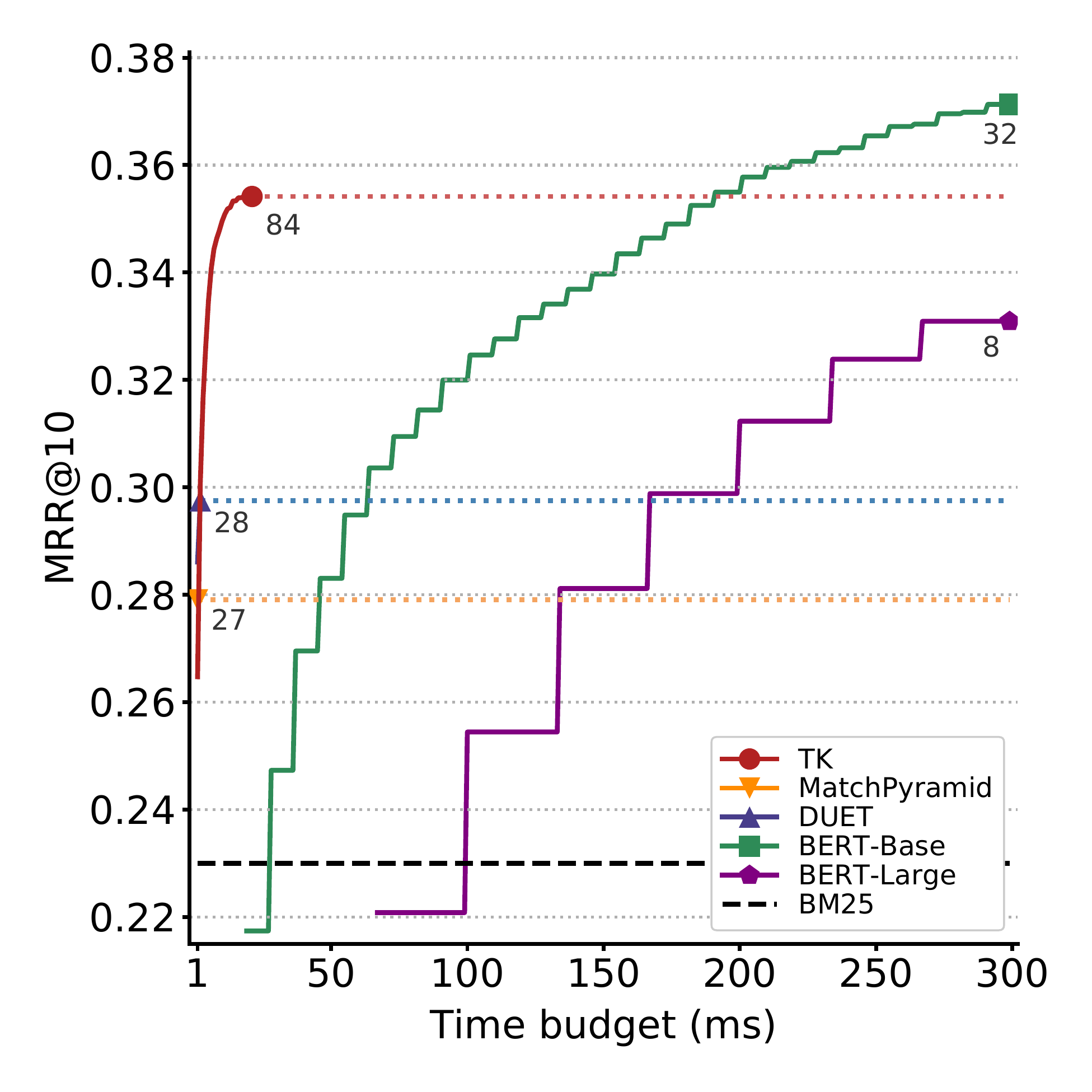}
    \end{minipage}%
    \hfill
    \begin{minipage}{.30\textwidth}
    \includegraphics[clip,trim={0.5cm 0.5cm 0.5cm 0.5cm} ,width=\textwidth]{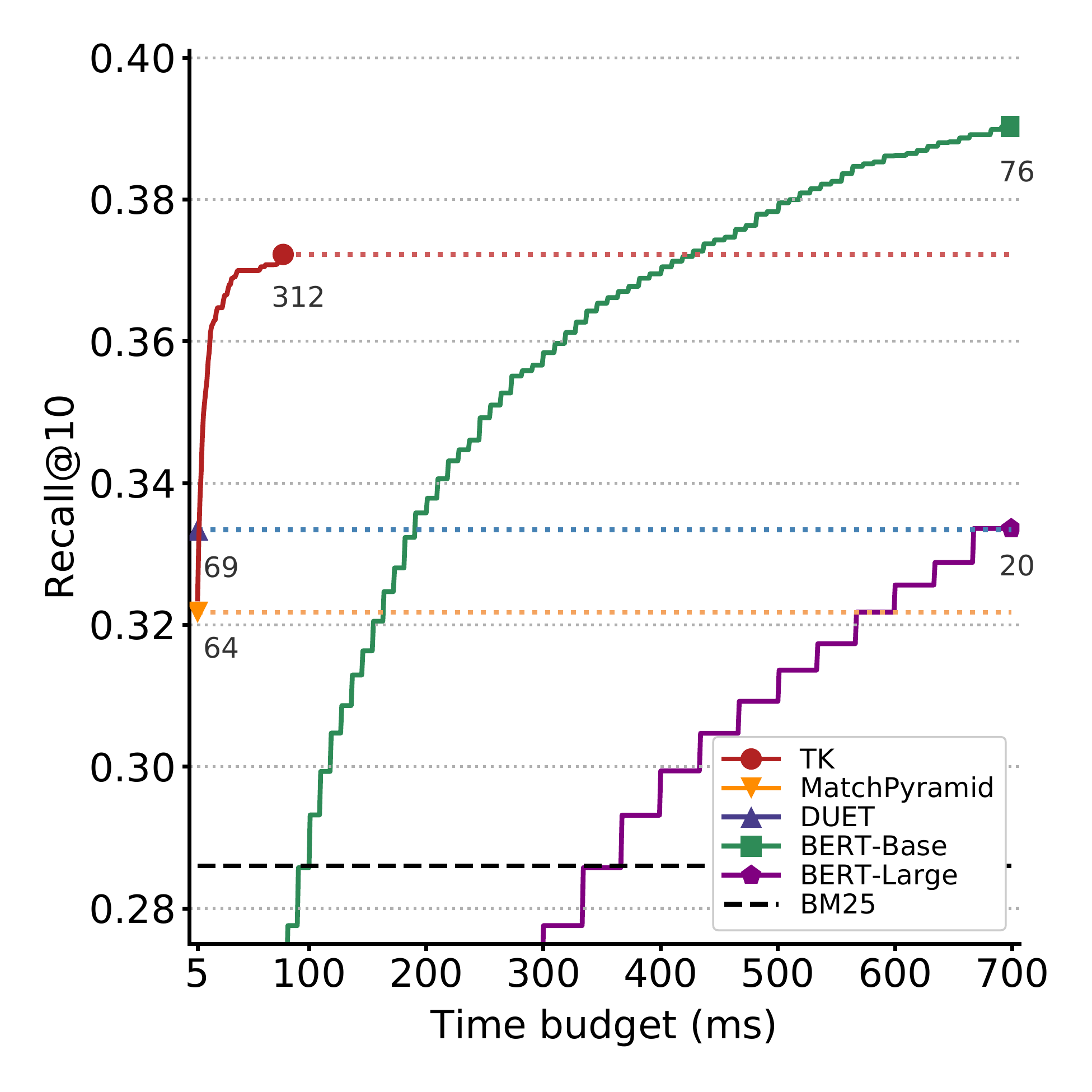}
    \end{minipage}%
    \hfill
    \begin{minipage}{.30\textwidth}
    \includegraphics[clip,trim={0.5cm 0.5cm 0.5cm 0.5cm} ,width=\textwidth]{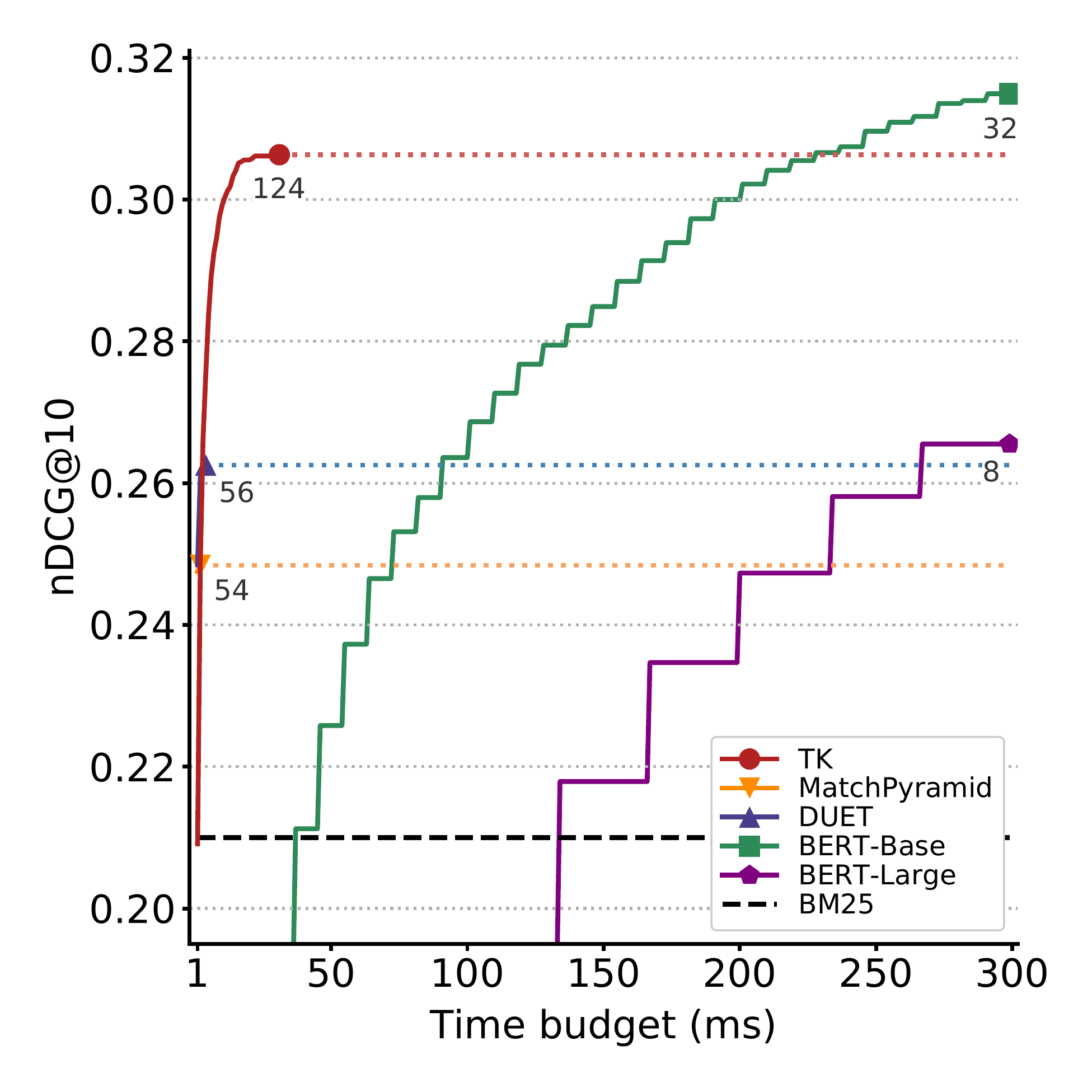}
    \end{minipage}
    
    \begin{minipage}{\textwidth}{\begin{center}\textbf{(c) TREC CAR}\end{center} }\end{minipage}%
    \centering
    \centering
    \caption{Time-budget analysis: We show the effectiveness (y-axis) of each model on the respective validation set by selecting the maximum number of documents to re-rank in the available time limit (x-axis). The marker indicates the best possible result; the dotted line indicates that additional time does not yield better results; the number of re-ranked documents is indicated for each best result.}
    \label{fig:budget_plots}
\end{figure*}

\vspace{-0.1cm}
\section{Results}
\label{sec:results}
\vspace{-0.1cm}
We first present the highest achievable effectiveness results for our evaluated models without any time limit (Section \ref{sec:unconstrained_eval}). Then, we present a novel time-budget evaluation, which is based on the realistic assumption that we trade effectiveness for efficiency and that users expect fast search results (Section \ref{sec:time_budget}). 
\vspace{-0.1cm}
\subsection{Effectiveness Evaluation}
\label{sec:unconstrained_eval}
\vspace{-0.1cm}
In Table \ref{tab:all_results} we show the highest achievable effectiveness results per model, without a time-constraint. The first section contains the traditional baselines; the second contains the neural re-ranking baselines; in the third section we report the results of our TK model with three different Transformer layer settings. Beside the effectiveness measures (MRR@10, Recall@10, nDCG@10 -- higher is better) we also report the best re-ranking depth per model, tuned on the validation set. We view this tuned parameter as a good indicator of a model's robustness on a collection and a useful analytical tool for the degree of difficulty of a collection. To incorporate the efficiency in the analysis, we report the average number of documents each model is able to re-rank per millisecond.

Of the three collections, the neural re-ranking models deliver the best results on MSMARCO-Passage, both in terms of effectiveness and re-ranking depth. Even simple neural baselines like KNRM and PACRR show significant effectiveness increases to the initial ranking baselines. TK and BERT show very strong results, especially on Recall@10, as they are able to incorporate almost all 1000 documents per query that we evaluated. Naturally, the more documents are re-ranked the higher is the potential for the Recall, as the Recall@10 is bound by the Recall of the first stage at the re-ranking depth. The more Transformer layers we use to contextualize embeddings in TK, the better the effectiveness becomes across all three measures, however, the differences are small. 

TK performs similar on the MSMARCO-Document collection as on the passage collection, however in general the results are closer together. The classic baselines are stronger as well as the non-BERT neural models. On the other hand both BERT results are reduced in comparison to their passage result. Overall, all neural models stagnate or reduce their effectiveness with a deeper re-ranking depth.

While BERT shows strong results on TREC CAR, it is especially challenging for non-BERT approaches. Except for MatchPyramid and DUET all non-BERT baselines fail to improve MRR@10 and nDCG@10 significantly. TK is the first non-BERT model offering strong improvements over the initial ranking baselines.

In conclusion, given unlimited time, BERT is the best re-ranking model, followed by TK, across all three collections. However, taking into consideration the average time a model spends on re-ranking a document it becomes apparent that utilizing BERT to its best re-ranking depth is prohibitively slow for most search applications.  
\input{contents/interpretability-figure.tex}

\subsection{Time-Budget Evaluation}
\label{sec:time_budget}
\vspace{-0.1cm}
Now we focus on the relationship between efficiency and effectiveness. It would be easy -- yet unfair -- to discard BERT as unfeasibly slow, based on the assumption that one always has to re-rank a thousand documents. Similarly, it would be unwise to solely judge the neural re-ranking models based on their unconstrained effectiveness results from the previous section. 

We take a fine-grained approach to evaluate efficiency and effectiveness together, by starting from the assumption that search applications set a time-budget for various stages of the retrieval pipeline. Neural re-ranking models are a part of a larger system and therefore have to adhere to a maximum time-budget. We use the re-ranking depth to control the time spent by each model. We believe this to be a fair comparison, as we give each model the same time. The timings we measured exclude any pre-processing and solely focus on the time spent computing the scores on the GPU. We evaluated each model once and then pruned the documents, based on their first stage rank, to obtain results for every re-ranking depth. We average the \textit{documents per millisecond} metric over all validation runs during the training, to obtain a noise reduced value. In a pilot study, we ensured that different validation batch sizes do not contradict the analytical results presented here. 

Figure \ref{fig:budget_plots} shows the time-budget aware results for every collection on MRR@10, Recall@10, and nDCG@10. The x-axes show the available time in milliseconds (up to 300ms for MRR and nDCG; 600 ms for Recall) and each y-axis represents the effectiveness results. We selected TK with 2 layers as a good compromise between effectiveness and speed. Additionally, we report the results for both BERT sizes and the two best non-BERT neural baselines. 

On the MSMARCO-Passage collection all fast models reach large re-ranking depths, except for BERT, where the base version only reaches 32 documents after 300 ms and the large version is only able to process 8 documents. TK is the best choice, after the first few ms of noise up to 190ms for MRR@10, 600 ms for Recall@10 and 300 ms for nDCG@10. BERT-Base overtakes the other neural baselines in around half the time it needs to be better than TK. If we choose a generous time-budget of 100 ms, TK's MRR@10 is 10\% higher, Recall@10 is 40\% higher, and nDCG@10 is 19\% higher than BERT-Base. BERT-Base can only re-rank 10 documents in 100 ms, leaving it at the same Recall@10 as BM25. Even at 250 ms, when TK finished all thousand documents it has a 12 \% higher Recall@10 than BERT-Base and is 9\% above CONV-KNRM. 

The MSMARCO-Document and TREC CAR collection are more challenging for non-BERT models, as their best re-ranking depth is shallow and more time would not yield better results. In Figure \ref{fig:budget_plots} this is shown as the colored dotted line. However, this does not change the time BERT needs to cross the best result of the other models. TK is the best choice for MSMARCO-Document up to 200ms for MRR@10, 550 ms for Recall@10 and 260 ms for nDCG@10. For TREC CAR, TK yields the highest effectiveness for a time-budget up to 200ms for MRR@10, 480 ms for Recall@10 and 250 ms for nDCG@10. If we again apply a time-budget of 100 ms to TREC CAR, we observe that TK's MRR@10 is 12\% higher, Recall@10 is 31\% higher, and nDCG@10 is 17\% higher than BERT-Base.

\subsection{Query Analysis}
\label{sec:query_analysis}
\vspace{-0.2cm}

Following the presentation of the collection results, we now analyze the query results in more detail. Our aim is to understand TK's quality on different query types or information needs and how TK compares to BM25 and BERT-Base. We cluster MSMARCO-Passage validation queries based on their mean contextualized embedding of TK with k-means. We set $k=30$, as we found it to be an appropriate choice with the elbow method. In Table \ref{tab:query_clusters} we show a selection of those clusters and their median rank of the first relevant passage for each model, as a robust measure inspired by MRR. Additionally, we report the number of queries in each cluster (\#~Q). We manually assigned an information need or \textit{type of query} summary to each cluster. In practice we observed most clusters to be unambiguous in their assignment, except for a few outliers per cluster. To keep the analysis simple, we do not place a time-constraint on the models.

The MSMARCO collections were created for question answering tasks and the queries reflect that. Only a minority of queries represents plain keyword queries -- a type for which BM25 provides good results. Natural language question queries are particularly well suited for neural re-ranking models, especially when users ask for a definition or clarification ("what is") with two or more words. Here, both TK and BERT improve substantially over BM25, while BERT performs slightly better than TK. BERT has an advantage over TK on complex queries with more than 8 words, which suggest the language modelling in BERT is more useful. 

\begin{table}
    \centering
    \vspace{-0.2cm}
    \caption{Comparing the median ranks of the first relevant document for BM25, TK, and BERT-Base on selected query clusters on the MSMARCO-Passage validation set}
    \label{tab:query_clusters}
    \begin{tabular}{lr!{\color{lightgray}\vrule}rrr}
       \toprule
       \multirow{2}{*}{\textbf{Information need}} & 
       \multirow{2}{*}{\textbf{\# Q}} &
       \multicolumn{3}{c}{\textbf{Median Rank}} \\
       && BM25 & TK & BERT \\ \midrule
       \textit{company} phone number  & 115 & 2  & 2  &  2  \\ 
       celebrity/movie facts & 224 & 8 & 4  &  3  \\ 
       money: cost/salary/net worth & 210 & 7 & 4  &  3  \\ 
       plain keyword(s) of diff. topics & 224 & 6 & 4  & 2   \\ 
       how long is \textit{something}  & 196 & 41 & 12  &  5  \\ 
       long question for a single number & 326 & 15 & 7  &  5  \\ 
       what is/are \textit{1 word}  & 321 & 15 & 3  & 3   \\ 
       what is/are \textit{2+ words}  & 279 & 20 & 5  &  3  \\ 
       meaning/definition of \textit{1 word}  & 155 & 13 & 3  &  3  \\ 
       meaning/definition of \textit{2+ words} & 208 & 23 & 4  &  3  \\ 
        symptoms of \textit{diseases}  & 58 & 59 & 9  & 5   \\ 
        benefits/effects of \textit{prescriptions}  & 106 & 13 & 4  & 4   \\ 
        \midrule
        \textbf{All queries} (with $\geq$ 1 relevant) & \textbf{6058} & \textbf{13} & \textbf{5} & \textbf{ 3} \\
        \bottomrule
    \vspace{-0.3cm}
    \end{tabular}
\end{table}

\section{Interpretability}
\label{sec:interpretability}
\vspace{-0.1cm}
We now highlight the interpretation capabilities of the TK model with a qualitative example. This analysis is enabled by TK's architecture, which first contextualizes query and document sequences independently and then uses only a single interaction value per term pair, which is scored via soft-histogram kernels. This allows us to accurately represent the scoring process of the model in our analysis. In contrast BERT-based re-rankers cannot be analyzed in this way, as they have no clear single point of interaction and a much more complex scoring mechanism.

We focus on the following scenario: a user would like to know why the neural model replaced the first result (a non-relevant document) of the first stage ranking with the actual relevant document. For this, we offer a side-by-side comparison view of two documents. Figure \ref{fig:interpretability_2col} shows the comparison of two documents for the query \textit{``androgen receptor define''}. On the left side is a document judged as relevant, which is placed on the first position by TK. On the right side is the top BM25 document, which is not the correct answer and only partially relevant to the query -- TK moved it to a lower position. 

We show each document with its full-text and a selection of temporary results of TK. We aim to identify and highlight the differences that result in different ranking scores. We color words according to their closest affiliation with a kernel. An important fact to consider is the soft-matching nature of the kernels: A term is counted in more than one kernel at a time. For example, this explains the difference in kernel $\mu=1$, even though no word is closest associated with that kernel and therefore we omitted a color.

From the highlighted kernel scores  (${s^{k}_{\text{log}}}$) it is apparent that the left document has more stronger matches than the right one, leading to higher scores. If we look at the corresponding colored words we observe that the sentence containing the definition in the left is most relevant to the query: \textit{{\color[RGB]{0,151,20}\uline{The}} {\color[RGB]{0,151,20}\uline{androgen}} {\color[RGB]{190,60,60}\uuline{receptor}} {\color[RGB]{0,151,20}\uline{(}} {\color[RGB]{0,151,20}\uline{AR}} {\color[RGB]{0,151,20}\uline{)}} {\color[RGB]{0,151,20}\uline{,}} {\color[RGB]{0,151,20}\uline{also}} {\color[RGB]{0,151,20}\uline{known}} {\color[RGB]{0,151,20}\uline{as}} {\color[RGB]{0,151,20}\uline{NR3C4}} {\color[RGB]{0,151,20}\uline{(}} {\color[RGB]{145,145,149}nuclear} {\color[RGB]{0,151,20}\uline{receptor}} {\color[RGB]{0,151,20}\uline{subfamily}}}. Even though TK does not contain a mechanism for strictly categorizing a region as relevant, it does so indirectly by strongly matching almost every term in this region. Of particular interest to us is the fact that the contextualization of TK learns to match the query term \textit{``define''} with words and phrases that make up a definition: \textit{``also known as''}, \textit{``subfamily''}, \textit{``is a type''} as well as the parentheses. This exceeds simple synonym mapping, suggesting once more the importance of training contextualized and relevance specific encoding models. 

This analysis demonstrates the potential for future work on keyword based search. When a collection is not queried with natural language questions, but only keywords, one could expand such keyword queries with terms like \textit{``definition''} or \textit{``meaning''} both during training and inference of neural models, to promote documents closer related to the core of the information need. 

%% file: contents/interpretability-figure.tex
\begin{figure*}[!t]
    \centering
    \noindent
    \begin{minipage}{\textwidth}
    \begin{center}
        {\color[RGB]{76, 76, 90}Query (Id:2)} \textbf{androgen receptor define}
    \end{center}
    \vspace{0.05cm}
    \end{minipage} 
    \begin{minipage}{.5\textwidth}
    \centering
      Rank: TK \raisebox{.5pt}{\textcircled{\raisebox{-.85pt} {1}}}, BM25 \raisebox{.5pt}{\textcircled{\raisebox{-.85pt} {9}}} \textbf{(judged as relevant}, Id: 4339068) 
      \newline
    \end{minipage}
    \begin{minipage}{.5\textwidth}
    \centering
      Rank: TK \raisebox{.5pt}{\textcircled{\raisebox{-.85pt} {8}}}, BM25 \raisebox{.5pt}{\textcircled{\raisebox{-.85pt} {1}}} (not relevant, Id: 1782337)
      \newline
    \end{minipage}
    \begin{minipage}[t]{.35\textwidth}%
    \vspace{-0.22cm}
{\color[RGB]{0,151,20}\uline{The}} {\color[RGB]{0,151,20}\uline{androgen}} {\color[RGB]{190,60,60}\uuline{receptor}} {\color[RGB]{0,151,20}\uline{(}} {\color[RGB]{0,151,20}\uline{AR}} {\color[RGB]{0,151,20}\uline{)}} {\color[RGB]{0,151,20}\uline{,}} {\color[RGB]{0,151,20}\uline{also}} {\color[RGB]{0,151,20}\uline{known}} {\color[RGB]{0,151,20}\uline{as}} {\color[RGB]{0,151,20}\uline{NR3C4}} {\color[RGB]{0,151,20}\uline{(}} {\color[RGB]{100,100,100}nuclear} {\color[RGB]{0,151,20}\uline{receptor}} {\color[RGB]{0,151,20}\uline{subfamily}} {\color[RGB]{100,100,100}3} {\color[RGB]{100,100,100},} {\color[RGB]{100,100,100}group} {\color[RGB]{100,100,100}C} {\color[RGB]{100,100,100},} {\color[RGB]{100,100,100}member} {\color[RGB]{100,100,100}4} {\color[RGB]{0,151,20}\uline{)}} {\color[RGB]{100,100,100},} {\color[RGB]{0,151,20}\uline{is}} {\color[RGB]{0,151,20}\uline{a}} {\color[RGB]{0,151,20}\uline{type}} {\color[RGB]{100,100,100}of} {\color[RGB]{100,100,100}nuclear} {\color[RGB]{0,151,20}\uline{receptor}} {\color[RGB]{100,100,100}that} {\color[RGB]{0,151,20}\uline{is}} {\color[RGB]{100,100,100}activated} {\color[RGB]{100,100,100}by} {\color[RGB]{100,100,100}binding} {\color[RGB]{100,100,100}either} {\color[RGB]{100,100,100}of} {\color[RGB]{100,100,100}the} {\color[RGB]{100,100,100}androgenic} {\color[RGB]{100,100,100}hormones} {\color[RGB]{100,100,100},} {\color[RGB]{100,100,100}testosterone} {\color[RGB]{100,100,100},} {\color[RGB]{0,151,20}\uline{or}} {\color[RGB]{100,100,100}dihydrotestosterone} {\color[RGB]{100,100,100}in} {\color[RGB]{100,100,100}the} {\color[RGB]{100,100,100}cytoplasm} {\color[RGB]{100,100,100}and} {\color[RGB]{100,100,100}then} {\color[RGB]{100,100,100}translocating} {\color[RGB]{0,151,20}\uline{into}} {\color[RGB]{100,100,100}the} {\color[RGB]{100,100,100}nucleus} {\color[RGB]{100,100,100}.} {\color[RGB]{0,151,20}\uline{in}} {\color[RGB]{100,100,100}some} {\color[RGB]{100,100,100}cell} {\color[RGB]{100,100,100}types} {\color[RGB]{100,100,100},} {\color[RGB]{100,100,100}testosterone} {\color[RGB]{100,100,100}interacts} {\color[RGB]{100,100,100}directly} {\color[RGB]{100,100,100}with} {\color[RGB]{100,100,100}androgen} {\color[RGB]{0,151,20}\uline{receptors}} {\color[RGB]{100,100,100},} {\color[RGB]{100,100,100}whereas} {\color[RGB]{100,100,100},} {\color[RGB]{100,100,100}in} {\color[RGB]{100,100,100}others} {\color[RGB]{100,100,100},} {\color[RGB]{100,100,100}testosterone} {\color[RGB]{100,100,100}is} {\color[RGB]{100,100,100}converted} {\color[RGB]{100,100,100}by} {\color[RGB]{100,100,100}5} {\color[RGB]{100,100,100}-} {\color[RGB]{100,100,100}alpha} {\color[RGB]{100,100,100}-} {\color[RGB]{100,100,100}reductase} {\color[RGB]{100,100,100}to} {\color[RGB]{100,100,100}dihydrotestosterone} {\color[RGB]{100,100,100},} {\color[RGB]{100,100,100}an} {\color[RGB]{100,100,100}even} {\color[RGB]{100,100,100}more} {\color[RGB]{100,100,100}potent} {\color[RGB]{100,100,100}agonist} {\color[RGB]{100,100,100}for} {\color[RGB]{100,100,100}androgen} {\color[RGB]{0,151,20}\uline{receptor}} {\color[RGB]{100,100,100}activation} {\color[RGB]{100,100,100}.}
\end{minipage}
    \hfill\begin{minipage}[t]{.14\textwidth}%
    \centering
    \begin{tabular}[t]{lr}
         \textbf{$\boldsymbol\mu_{k}$} & \textbf{$\boldsymbol{s^{k}_{\text{log}}}$} \\
         1 & -3.1 \\
         {\color[RGB]{202, 70, 70}\uuline{0.9} }& {\color[RGB]{202, 70, 70}-0.6} \\
         {\color[RGB]{0,151,20}\uline{0.7}} & {\color[RGB]{0,151,20}2.3} \\
         0.5 & -1.6 \\
         0.3 & -3.3 \\
         Rest & -14.6\\
         \midrule
         \textbf{$\boldsymbol{s_{\text{log}}}$} & -11.6\\
         \textbf{$\boldsymbol{s_{\text{len}}}$} & 1.1\\
         \midrule
         $\boldsymbol{s}$ & -10.6\\

    \end{tabular}
    \end{minipage}\hfill\vline\hfill
    \begin{minipage}[t]{.14\textwidth}%
    \centering
      \begin{tabular}[t]{lr}
         \textbf{$\boldsymbol\mu_{k}$} & \textbf{$\boldsymbol{s^{k}_{\text{log}}}$} \\
         1 & -5.0\\
         {\color[RGB]{202, 70, 70}\uuline{0.9}} & {\color[RGB]{202, 70, 70}-1.5}\\
         {\color[RGB]{0,151,20}\uline{0.7}} & {\color[RGB]{0,151,20} 1.9}\\
         0.5 & -1.3\\
         0.3 & -2.3\\
         Rest & -14.6\\
         \midrule
         \textbf{$\boldsymbol{s_{\text{log}}}$} & -12.8\\
         \textbf{$\boldsymbol{s_{\text{len}}}$} & 0.9\\
         \midrule
         $\boldsymbol{s}$ & -11.9\\
    \end{tabular}
    \end{minipage}
    \hfill\begin{minipage}[t]{.35\textwidth}%
    \vspace{-0.22cm}
{\color[RGB]{100,100,100}Enzalutamide} {\color[RGB]{0,151,20}\uline{is}} {\color[RGB]{0,151,20}\uline{an}} {\color[RGB]{0,151,20}\uline{androgen}} {\color[RGB]{0,151,20}\uline{receptor}} {\color[RGB]{100,100,100}inhibitor} {\color[RGB]{0,151,20}\uline{that}} {\color[RGB]{0,151,20}\uline{acts}} {\color[RGB]{100,100,100}on} {\color[RGB]{100,100,100}different} {\color[RGB]{100,100,100}steps} {\color[RGB]{100,100,100}in} {\color[RGB]{0,151,20}\uline{the}} {\color[RGB]{100,100,100}androgen} {\color[RGB]{0,151,20}\uline{receptor}} {\color[RGB]{100,100,100}signaling} {\color[RGB]{100,100,100}pathway} {\color[RGB]{100,100,100}.} {\color[RGB]{100,100,100}Enzalutamide} {\color[RGB]{100,100,100}has} {\color[RGB]{100,100,100}been} {\color[RGB]{0,151,20}\uline{shown}} {\color[RGB]{0,151,20}\uline{to}} {\color[RGB]{100,100,100}competitively} {\color[RGB]{100,100,100}inhibit} {\color[RGB]{100,100,100}androgen} {\color[RGB]{0,151,20}\uline{binding}} {\color[RGB]{0,151,20}\uline{to}} {\color[RGB]{100,100,100}androgen} {\color[RGB]{0,151,20}\uline{receptors}} {\color[RGB]{100,100,100}and} {\color[RGB]{100,100,100}inhibit} {\color[RGB]{100,100,100}androgen} {\color[RGB]{0,151,20}\uline{receptor}} {\color[RGB]{100,100,100}nuclear} {\color[RGB]{100,100,100}translocation} {\color[RGB]{100,100,100}and} {\color[RGB]{100,100,100}interaction} {\color[RGB]{100,100,100}with} {\color[RGB]{100,100,100}DNA} {\color[RGB]{100,100,100}.}
\end{minipage} 
    \caption{TK's scoring results of two MSMARCO-Passage documents: We highlight two close similarity kernels (0.9 \& 0.7). In the text words are colored and underlined if they are closest to the center of the kernel. Individual kernel results (model weights included) are displayed in the middle for each document.}    \label{fig:interpretability_2col}
\end{figure*}
\vspace{-0.4cm}

%% file: contents/6-conclusion.tex
\vspace{-0.1cm}
\section{Conclusion}

The work in this paper is based on the assumption that search tasks are time-constrained and neural re-ranking models have to fulfil this requirement to be deployable as part of user-facing search engines. To address this, we proposed TK: an interpretable ad-hoc neural re-ranking model with a very strong efficiency-effectiveness ratio. We introduced a realistic time-budget aware evaluation. Models are allowed to re-rank as many documents as they can within the given time-budget. This evaluation shows how the TK model is the overall best choice for a time-budget under 200 ms per query for precision based measures and 400 ms per query for recall. In addition -- to not just propose a black-box model without insight -- we provided a fine granular query analysis showing the different strengths of TK on various query types and we illustrated how TK can be analyzed and interpreted. TK makes it possible to obtain competitive neural re-ranking results with a limited time-budget.
\paragraph{\textbf{Acknowledgements}} This work has received funding from the European Union's Horizon 2020 research and innovation program under grant agreement No 822670.